\pdfoutput=1

\documentclass[11pt, conference, compsocconf]{IEEEtran}
\usepackage{color}

\usepackage[utf8x]{inputenc}
\usepackage[pdftex]{hyperref}
\usepackage{url}

\usepackage{draftcopy}

\newif\ifwaterzero
\waterzerotrue
\ifwaterzero
\usepackage{type1cm}
\usepackage{eso-pic}

\fi

\newif\ifwaterone
\ifwaterone
\usepackage{draftwatermark}

\fi

\usepackage{ieeefig}
\usepackage[pdftex]{thumbpdf}

\hypersetup{
    pdfauthor         = {Janis Danisevskis, Michael Peter, Jan Nordholz},
    pdftitle          ={Minimizing Event-Handling Latencies in Secure Virtual Machines},
    hypertexnames     = true,
    hyperfigures      = true,
    pagebackref       = true,
    colorlinks        = true,
    linkcolor         = blue}

\newif\iftodo
\todotrue

\newcommand{\note}[1]{}

\newlength{\benchfigurewidth}
\newlength{\figurewidth}
\begin{document}

\title{Minimizing Event-Handling Latencies in Secure Virtual Machines}

\author{
\IEEEauthorblockN{Janis Danisevskis, Michael Peter, Jan Nordholz}
\IEEEauthorblockA{Technische Universit\"at Berlin\\
       Deutsche Telekom Laboratories\\
       Security in Telecommunications\\
       \{janis, peter, jnordholz\}@sec.t-labs.tu-berlin.de}
}

\date{03 Feburary 2012}


\maketitle


\begin{abstract}

  Virtualization, after having found widespread adoption in the server
  and desktop arena, is poised to change the architecture of embedded
  systems as well. The benefits afforded by virtualization --- enhanced
  isolation, manageability, flexibility, and security --- could be
  instrumental for developers of embedded systems as an answer to
  the rampant increase in complexity.

  While mature desktop and server solutions exist, they cannot be
  easily reused on embedded systems because of markedly different
  requirements.  
  Unfortunately, optimizations aimed at throughput, important for servers,
  often compromise on aspects like predictable real-time behavior, which
  are crucial to many embedded systems.
  In a similar vein, the
  requirements for small trusted computing bases, lightweight inter-VM
  communication, and small footprints are often not accommodated. 
  This observation suggests that virtual machines for embedded systems
  should be constructed from scratch with particular attention paid to the 
  specific requirements.

  In this paper, we set out with a virtual machine designed for
  security-conscious workloads and describe the steps necessary to
  achieve good event-handling latencies. That evolution is possible
  because the underlying microkernel is well suited to satisfy
  real-time requirements.  As the guest system we chose Linux with the
  {\tt PREEMPT\_RT} configuration, which itself was developed in an
  effort to bring down event-handling latencies in a general purpose
  system. Our results indicate that the increase of event-handling
  latencies of a guest running in a virtual machine does not, compared
  to native execution, exceed a factor of two.

\end{abstract}


\IEEEpeerreviewmaketitle

\section{Introduction}

\label{sect:intro}

Historically, innovations first took place in the desktop and server
market before they trickled down to embedded systems some years later.
Although the gap in capabilities is likely to persist, high-end
embedded devices more and more resemble desktop systems in term of
hardware (compute power, memory sizes) and software (general purpose
operating systems, open system architecture).
Assuming that this follow-suit trend is to continue, it should not be
too long before virtualization gains traction in embedded devices.
Before dwelling on real-time characteristics, we would like to briefly
recite the arguments for virtualization in embedded devices.

\subsection{The Case for Virtualization in Embedded Systems}

\emph{Development support.}  Tight time-to-market times can only be
achieved though pervasive software reuse.  Despite all efforts,
dependencies between software components are often intricate,
requiring substantial development effort when changes are needed.
Moreover, quite a few applications hinge on particular operating
systems, or even specific versions of them. Virtualization obviates the
need to settle for one operating system, but instead allows to run
multiple of them in parallel. The easy reuse of tried-and-tested
components will cut down on the development effort thereby reducing
costs and shortening the time-to-market.

\emph{Platform migration.}  In the past, more powerful devices
automatically translated into better functionality. The recent
tendency to multicore processors makes that benefit increasingly
harder to reap as only multithreaded software benefits from the
presence of additional cores. Virtualization opens up the opportunity
to eschew the daunting task of porting single-core software stacks,
instead reusing it outright. Generally, the introduction of
virtualization into the system stack requires only the hypervisor to
be ported to new devices; the higher layers need only small adoptions
or none at all.

\emph{Security.} Traditionally, security from non-physical attacks was
not a major concern for embedded systems.  Even if a device contained
defects, these were difficult to exploit as the usual attack vector
--- network connectivity --- was unheard-of.  That has changed
profoundly in the recent past. High-end embedded systems more and more
resemble desktop systems regarding computing capabilities, usage
scenarios, and connectivity. The applications they accommodate share
many traits with those found on desktops, security vulnerabilities
included. Security-wise, the calm past has given way to a frantic
struggle between attackers raking about for exploitable deficiencies
and defenders hurrying for closing them promptly.  While  security
threats against desktops are severe, they hold the potential to be
devastating for embedded systems. Unlike the stationary cousins,
embedded systems often are restricted in their means to repulse
malware. Installing an new version of an anti-virus solution is just
not on the table. Given the gravity of the situation and the
inevitability of its occurence, it is urgent to tighten the defences.
Virtualization may provide the isolation needed to contain an
attacker, thus offering deeper defense. The feasibility of this
approach has been borne out by contemporary gaming consoles, where
hypervisor-based security architectures have proven sufficiently capable
of denying ultimate control to flocks of highly motivated attackers.

\emph{Consolidation.} Today it is not uncommon to physically separate
subsystems to ensure non-interference. The downside is that the number
of control units grows which raises concerns as to the bill of
material, weight of wiring, and reliability in the face of numerous
physical connections, which are vulnerable to aging.

\emph{License separation.} Platform vendors are often concerned about
their system architecture not to be revealed to competitors.  For that
reason, they do not like the publication of, say, device drivers as
these may give hints at the underlying hardware. This attitude is at
odds with some important open source operating systems, which require
source code access for device drivers. The tension can be eased if
software with intellectual property rights restrictions is placed in a VM where
it is not subject to licenses requiring source access.

\emph{Partial Validation.} If the interaction between components
cannot be characterized precisely, then a tightly integrated system
has to be revalidated as a whole whenever any single component is
changed. Conversely, as long as component interaction is limited to
well-defined interfaces (including temporal aspects), it is
sufficient to check whether a component adheres to the interface
specification. Changes of a component then would not require a system
validation.

\emph{Remote Management.}  The growing device complexity makes it ever
more likely that some unexpected situation arises after the device is
shipped.  Therefor it would be desirable to remotely assess the
situation and take remedial measures. It the best case, the issue is
fixed by a remotely initiated update before the user even notices that
something was about to break.
For all its virtues, remote management cuts both ways. If not properly
secured, a remote management interface is a downright invitation for
attackers to take over the device. As such, it must be ensured that
only properly authorized requests are honored. There should be a
\emph{secure anchor} in the platform, which cannot be even disabled if
an attacker has already succeeded in penetrating some parts of the
system.

\subsection{Challenges for Embedded Virtualization}

Embedded devices differ from their desktop and server cousins in that
they are often charged with tasks that require interacting with their
environment in real-time. The transition from special-purpose
real-time operating systems to general purpose operating systems like
Linux poses enough problems in itself in that respect. Virtualization
requires modifications deep in the software stack. In order to hurt
real-time performance not too much the following three problems have to be
taken into account.

First, virtualization support in current processors comes in the shape
of additional processor privilege levels\footnote{The details differ
  though.  Some instruction set architectures added to their existing
  privileges less privileged modes (x86 guest mode (AMD) and non-root
  mode (Intel)) while other went for more privileged modes (PPC
  hypervisor mode).}. Not only have incoming event now to traverse
more modes (CPUs deliver events normally to the most privileged
mode), the state transitions also take longer due to the huger state involved.
With current hardware, these additional expensive transitions in
timing-critical paths are inevitable and will lead to increases in
event-handling latencies.

Second, the relative cost of individual operation changes, which
requires a redesign of timing critical paths. For example, on a
non-virtualized platform arming a timer takes a trap into the
operating system kernel and a device access. In a virtual machine, a
guest is not allowed to access the physical timer directly because of
contention from other guests. Attempts to access the device trap into
the hypervisor, which takes care of multiplexing the physical timer.
Unfortunately, this costly operation is in critical event handling
paths. An alternative is to let the hypervisor arm the timer
speculatively when it injects an event. The guest OS can then resume
the event handler process without the delay incurred during setting
the (virtual) timer.

Lastly, the hypervisor has to be aware of the internal operations of a
VM. Assume a process in a VM that runs with the highest priority and
shall complete execution as fast as possible. Further assume that,
during the execution of that process, an IRQ arrives. The hypervisor
takes over control and dispatches the event to the virtual machine
monitor\footnote{We assume that the VMM runs as a user-level task.},
which in turn injects it into the VM. Only then the decision is taken
that the event does not bear on the scheduling configuration and the
high priority task is resumed. At that point an appreciable amount of
time has elapsed due to the privilege changes and context switches .
Had the hypervisor or the VMM had the information that the IRQ does
not affect the current scheduling decision in the VM, either one could
have abstained from forwarding the event and instead directly resumed
the interrupted activity. This short-circuiting, though, is only
possible if it is known that a high priority process is executing.
Otherwise, forwarding the event is the correct action. As the
architecturally defined VM interface does not convey this kind of
information, additional interfaces are needed to bridge the {\emph
  semantic gaps} that opens up between hypervisor, VMM, and guest VM.

\subsection{Goal and Outline}

Our goal was to build a system with the following properties:

\begin{enumerate}
\item The {\emph {trusted computing base}} --- the set of all hardware,
  firmware, and software components that are critical security and
  safety of a system --- shall be small. It is well accepted that a
  reduction in complexity is instrumental in building reliable
  systems.
\item Given the scarcity of native applications for security-conscious
  environments, the reuse of legacy software shall be easily possible.
  As virtualization is by definition compatible with a huge body of
  software and exhibits good performance characteristics, the system
  shall support multiple virtual machines.
\item One of this virtual machines shall be able to pick up promptly
  on incoming events. The hosted guest shall see event-handling
  latencies that are within a reasonable range of that seen on
  physical hardware.
\end{enumerate}

The contribution of this paper is twofold: first, we
present a  software stack that takes into account  both real-time and security requirements. We assess the impact of
this arrangement on real-time workloads running in a VM. 

In the second part, we investigate how real-time behavior can be improved
if the guest provides hints regarding its real-time operations.
Although such hints require guest modifications, we found that these
modifications can be kept to a minimum and improve significantly on
the real-time performance.

\section{Background}
\label{sect:background}

Requirements regarding security, fault-tolerance, and real-time
performance can only be met if the lowest layer in the software stack,
the operating system kernel, provides proper support. One long-standing
quarrel on this issue is whether monolithic kernels can be adequately
evolved or a shift towards microkernels is the better choice. While a
huge part of the OS community recognizes the principal superiority of
the microkernel design, many point out that monolithic kernels can
catch up and retain their performance advantage.

\subsection{Microkernels}

Operating system kernels are a critical component in the software
stack. Kernels find themselves in the situation that they have to
guarantee safety and security for \emph{all} applications running on
top of them, yet cater to specific needs of single applications.
Applications often have very specific requirements, which are not
easily conveyed through the syscall interface. An operating system
would like to hand over the resource management to applications as
that would allow to apply arbitrary policies without adversely
affecting non-participating parties.

A microkernel is an operating system kernel that contains only those
mechanisms needed to implement a complete operating system on top.
Typically these mechanisms include address space management, scheduling,
and message passing. Other functionality such as device drivers,
protocol stacks, and file systems is provided by user-level servers. The
upside of this approach is that resources can be selectively allocated
to user-level tasks, which then can manage them at their discretion.

While hailed as the future of operating systems, microkernels have not
supplanted monolithic kernels across the board. Although many
experimental microkernels demonstrated the potential benefits, neither
one saw enough take-up, which would have been necessary for a self-sustaining
software ecosystem. Furthermore, systems built on monolithic kernels
kept their performance edge and had some of their deficiences
alleviated.  For example, the often raised point of lacking
flexibility was answered by the introduction of kernel modules.
The lack of success on desktops and servers notwithstanding,
microkernels made their way into systems that care about security and
scheduling.

\paragraph{Scheduling}

When a computation involving updates of multiple variables can be
preempted, further synchronization mechanisms have to be employed to
ensure consistency. While lockless schemes such as
RCU\cite{citeulike:402420} and lock-free
synchronization\cite{Herlihy:1993:MIH:161468.161469} have an edge over
classical lock-based synchronization in certain cases, the latter is
easier to understand and thus prevalent.

Retrofitting preemptibility into
monolithic kernels by adding locks to data structures is tedious
because assumptions as to atomic execution are usually deeply
entrenched and non-obvious. Deadlock situations may arise if the lock
grabbing sequence does not rule out circular dependencies. If the
objects to be protected come into being and perish at a high rate,
coming up with a suitable lock ordering is far from trivial. The
situation is aggravated if long running operations shed held locks
intermittently and reacquire them before resuming.  Such a courtesy
may be imperative, though, if operations with tighter timing
requirements require prompt access. 
Furthermore, for most real systems, event-handling latencies are not
an overriding priority and thus must not hurt throughput performance
too much. As a consequence, the number of locks in often-used code paths
has to be limited, which may result in longer lock-protected paths.
The problem was encountered by developers of the \emph{PREEMPT\_RT}
flavor of the Linux kernel. Up to 15\% performance
degradation\cite{preempt:rt:overhead} was enough to stir opposition
against mainline inclusion.

Given the difficulties of directly improving on the preemptibility of
an existing kernel, it may be expedient to absolve it from scheduling
and assign that task to a real-time scheduler. For that purpose, a
shim is interposed between the old kernel and the hardware, taking
over the primary responsiblity for scheduling. The old kernel is -
from a scheduling point of view - relegated to a real-time tasks.  As
such, it can stop the flow of events from the shim to it.
However, it cannot prevent events arriving at the shim and triggering
scheduling decisions there.

The idea has gained some momentum, both in open source and commercial
project. One of the first projects along this architecture was
RTLinux\cite{RTLinux}.  RTAI\cite{1199229} and its successor, the
Xenomai project followed suit. The downside of this approach is that
the shim only takes over scheduling but does not assume control over
memory management. Memory management still lies with the host kernel,
in these cases Linux. Real-time tasks are loaded into the Linux kernel
and become subject to the real-time scheduling after initialization.
The common residency in kernel space means that crashes are not
contained in address spaces. Xenomai tries to mitigate the issue by
optionally placing real-time applications in address spaces.  While
faults of other real-time tasks become survivable, crashes in the
Linux kernel are still fatal.

The next step is not only to confer scheduling to the shim but also
hand the responsibility for memory management completely over to it.
That would strengthen the surviveability of the system as more parts
are subject to address space isolation. The downside is that this
encapsulation carries the cost of privilege level transitions in
critical paths. Nonetheless, many system have adopted that
architecture, among them QNX Neutrino\cite{QNX}, PikeOS\cite{PikeOS},
and L4 \cite{1181568}.

\paragraph{Security}

Monolithic kernels have grown to a complexity were it becomes
intractable to reason about all possible interactions in the kernel. Given that
each defect may be used to gain control over the kernel, such a
situation is a security nightmare. In contrast, small kernels are more
amenable to thorough examination, which instills confidence regarding
their correctness. This scrutiny can be even taken as far as to have
them undergo formal verification\cite{Klein_EHACDEEKNSTW_09}.

A prerequisite for a small trusted computing base, a small kernel
alone is not sufficient, though. Rather it is necessary that
privileged user level servers cannot be tricked into exercising their
authority on behalf of a malevolent party. This issue --- widely known
as \emph{confused deputy problem} --- can be ascribed to the
separation of naming and privilege. 

\newif\ifseven

Capabilities facilitate the construction of system that honour the
\emph{pinciple of least authority}.  
Despite their undenied
advantages, capabilities have not found their way into mainstream
operating systems. The reason for that might well be that capabilities
are not compatible and thus cannot retrofitted into the APIs of
mainstream OSs. A recent effort by Watson at
al\cite{citeulike:7658009} shows promise to make deeper inroads but is
yet in early stages.

Researchers, who have not to take legacy APIs into consideration, were
more successful with creating conceptually clean systems.  The first
system that featured a small kernel and capability-based access
control, EROS\cite{Shapiro99eros:a}, was followed by more systems with
these
traits\cite{Lackorzynski:2009:TSC:1519130.1519135}\cite{Steinberg:2010:NMS:1755913.1755935}.

\subsection{Virtualization}

Microkernel-based systems are relatively novel under active
development and as such still in a state of flux. An steadily changing
kernel interface, though, is a difficult target for user-level
components.
Although native system frameworks have come into existence, they
cannot rival existing operating system in features, maturity, and
application availability.  To make up this deficiency, system
designers have chosen to reuse whole operation systems with virtualization
the most promising technique.

A virtual machine is an software environment which is capable of
executing applications together with their operating systems. A
virtual machine implements the same instruction-set architecture as
real hardware, a guest cannot distinguish whether it runs natively or
under virtualization. While this requirement can also be met by
performance-sapping emulation, virtual machines are required to
exhibit near-native performance. This requirement translates into the
huge majority of instructions being executed directly by the host
hardware. Complementing equivalence and performance is control.  The
virtualization layer always retains full control over the system.  To
that end the part of the VM in control --- the hypervisor ---
interposes on operations that impinge upon system resources such as
memory or IO devices. Instead of accessing the physical resource, the
guest is provided with virtual surrogates.
As encapsulation is an explicit virtualization goal, virtual
machines 
naturally lend themselves to serve  as a unit of isolation.

The advantages of virtualization come at a cost, though. Virtual
machines operate in special processor modes which incur substantial
overhead when entered or left. While also being a problem for systems
aiming at throughput, these world switching costs can add
substantially to the length of critical paths of real-time
applications if these were to run in VMs.

Apart from hardware-induced costs, software-related costs 
have to be considered, too.
Latencies incurred by executing non-preemptible code paths within the
layers implementing the VM will add to those caused by the OS kernel
in the VM. For this reason, it is not recommendable to use so-called
hosted VMMs, where an existing operating system is extended with
hypervisor functionaltiy. Bare-metal hypervisors can be implemented as
microkernels and as such have much less code in timing-critical execution
paths.

While bringing hardware and (direct) software costs under control
is necessary, it is not sufficient, though.  One severe problem with
virtualization architectures is that the machine interface does not
propagate high-level information, an issue also known as
\emph{semantic gap}.  For example, on bare hardware, there is no need
for an operating system to let the hardware know that there is a
highly prioritized real-time application runnable. In contrast, under
virtualization, a guest disabling interrupts achieves only that no
events arrive in \emph{its} VM. The system still accepts them and
queues them until they can be delivered. While the guest performs
correctly, it does not as fast as it could as interrupt arrival
entails costly world switches. If the lower layers were aware that
real-time tasks are active, they could disable (at least) some events
in order to expedite the guest execution. 
The information that a real-time task is
active is not easily derived by the host, thus it has to be signaled
by the guest.

\section{ Design}
\label{sect:design}

We will start the description of our design by covering the
hypervisor and those features with a bearing on real-time operations in VMs.
Thereafter we will elaborate on our virtual machine monitor.  The
chapter is wrapped up with the explanation of the changes we made to
the guest.

Before we set out to detail our design, we will take up an issue
that often gives rise to confusion.
While often used interchangeably, the terms \emph{hypervisor} and
\emph{virtual machine monitor} signify two distinctively different
components in our architecture. More confusingly, sometimes the
\emph{hypervisor} is used synonymously with \emph{microkernel}.

A hypervisor enforces isolation, that is it implements protection
domains, VMs being one flavor of them. For doing so, it needs to
exercise exclusive control over data structures that control isolation
such as page tables and entry vectors. This is only possible if it
runs in the the most privileged CPU execution mode\footnote{We only
  consider accessible execution modes, which leaves out the system
  management mode, which is arguably even more privileged.} as only
this mode has full control over critical system resources such as
control registers.

A microkernel applies the
principle of minimality to the portion of software running in the most
privileged execution mode. We understand microkernel and hypervisor
complementary. Hypervisor implies a function whereas microkernel
denotes a particular design principle. As such, a microkernel can
assume the role of a hypervisor, or conversely, a hypervisor can be
implemented as microkernel.

A virtual machine monitor provides VMs with functionality beyond CPU
and memory virtualization, which is the duty of the hypervisor.
Typically, the VMM supplies a VM with (virtual) devices and
coordinates its execution. Contrary to a hypervisor, which always
requires maximal CPU privileges, a VMM can be either part of the
kernel or implemented as a user task.

\subsection{Hypervisor}

Our architecture builds on
\emph{Fiasco.OC}\cite{Lackorzynski:2009:TSC:1519130.1519135}, a
state-of-the-art microkernel of L4 provenance.  As typical for
microkernels, it only provides basic functionality such as address
space construction, and inter-process communication (IPC), and
scheduling. All the remaining aspects such as device drivers and
protocol stacks, which are included in monolithic kernels, are
relegated into user level servers where they are subject to address
space containment. Under this arrangement, a device driver failure,
which is likely fatal in monolithic systems, becomes survivable.

The original Unix security model, which served as a role model for
many contemporary systems, does not allow for mandatory access control
(MAC). Each data owner may release data to arbitrary third parties at
his discretion without a system-wide imposed security policy being
able to prevent that. Efforts to add mandatory access control
mechanisms such as SELinux have proven difficult and are widely deemed
inpractical.  In contrast, Fiasco.OC features capability-based
access control\cite{Lackorzynski:2009:TSC:1519130.1519135}, which
highly expedient for the construction of systems that follow the
\emph{princple of least authority}.

Fiasco.OC can be used as an
hypervisor\cite{schild2009faithful}\cite{liebergeld2010secVM}. In line
with the microkernel philosophy, it only provides support for CPU and
memory virtualization and leaves the provisioning of virtual devices
to a user-level virtual machine monitor (VMM). It should be pointed
out that virtualization blends smoothly with the existing system.
Memory is provisioned into VMs the same way as it is into non-VM
protection domains.  Access to VMs is governed by the same capability
mechanism that controls access to any other object in the system.
Executing code in a VM is achieved through a thread and as such under
control of the Fiasco scheduler.

The object describing address spaces, the task, is also used to
provide the memory context for a VM. The only modification is that the
VM memory range covers the full 4GB on 32bit architectures whereas
tasks are limited to 3GB. Each virtual machine has an associated
controller task, which controls its execution thereby acting as a
virtual machine monitor (VMM).  Unlike tasks, VMs cannot host threads
directly. Instead, the thread is bound to the controller task and
migrates into the VM when execution shall there resume. To that end it
invokes a microkernel syscall with a capability to the VM object and a
full architectural VM state as arguments. This operation is reverted
when an external event arrives or a situation occurs that needs
attention. The last execution state of is then reported back.

From its inception, Fiasco was developed with real-time support in
mind. The kernel disables interrupts only for inavoidable atomic
operations (e.\,g. thread switching) and is thus highly preemptible.
Scheduling of threads is strictly based on static priorities.

Streamlined synchronous IPC was one of the defining characteristics of
the original L4 and later derived kernels. The reasoning was that
communication is mainly procedure-call-like and that asynchronous
event handling is so rare that kernel support is not warranted.
It turned out that the premise does not hold for operating system
kernels. Consequently, the kernel's threading model was augmented with
a virtual CPU mode \cite{lackorzynski2010vcpu}, which proved very
expedient for OS rehosting. The underlying principle is that control
transfer between two tasks\footnote{In L4Linux both the Linux kernel
  and Linux processes are implemented as L4 tasks.} only takes place
if there are no pending interrupts. That mirrors the behavior of
physical machines, where transitioning from kernel into user level is
and allowing interrupt delivery is usually an atomic operation.
Earlier designs lacking that feature could ignore pending events for a
whole time slice, which is unacceptable for real-time operations.  For
our work, the scope of this conditional operation was extended to the
\emph{switch into VM} operation.

\subsection{Virtual Machine Monitor}
\label{subsect:VMM}

Instead of starting from scratch, we opted to improve on the
\emph{Karma VMM}\cite{liebergeld2010secVM}.
We identified parts of Karma that might delay the delivery of events
into the guest and replaced them.

Our first modification concerned the timer handling. Originally, Karma
used Fiasco IPC timeouts as time source. For historic reasons,
Fiasco's timer granularity is rather coarse at 1000Hz. Clearly such a
clock source is inadequate to serve as a time source for a
high-resolution Linux timer. Since changing the Fiasco timer
infrastructure was beyond our scope, we decided to offer a
higher-resolution timer as an alternative. Our choice was fell on the
HPET.  Ownership over the device is given to one karma instance,
whereby it is allowed to program it directly and subscribe as a
recepient of its interrupts. Whenever the guest OS tries to program
its (virtual) timer, these accesses are intercepted. The VMM retrieves
the timeout to be set from the guest's execution state and arms the
HPET accordingly. When the time has elapsed, the HPET raises an
interrupt. Fiasco forwards it to the VMM, which, in turn, injects it
as a virtual interrupt into the VM under its control.

Since the VMM is a regular task it does not have direct device access.
Instead it uses services provided by infrastructure servers.  In many
cases, service invocation involves sending messages (interprocess
communication, IPC). Unfortunately, most of the currently used
infrastructure employs synchronous IPC to a certain degree.

However, a VMM does not fit well into this model. A VM may accommodate
multiple processes, each of which can potentially initiate
communication with other servers (by accessing virtual devices).
Since IPC response times are not guaranteed to be short\footnote{In
  fact, most services do not specify any timing parameters.}, it could
happen that any activity in the VMM came to an halt for a substantial
period of time. Any incoming event was delayed until an IPC reply
arrived.
To avoid these delays we employ a dedicated thread for external IPC.
The main thread still detects the request from the guest, but instead
of sending the IPC itself it leaves that task to the communication
thread. The important point in this construction is that the main thread
gets not tied up in an IPC and can promptly pick up on incoming events.

\section{Optimizations}

\begin{figure}
\resizebox{\columnwidth}{!}{\input{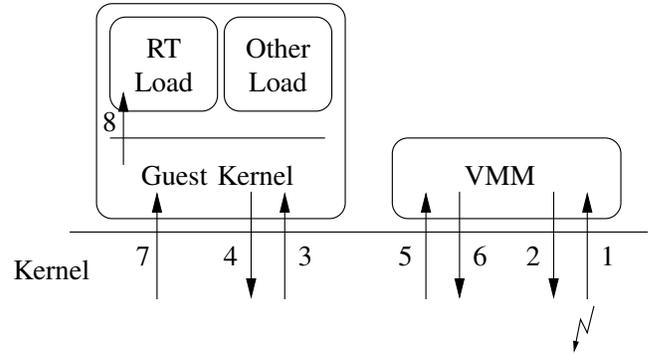}}
\caption{
Usual series of events upon arrival of a timer interrupt. \mbox{Steps~1-3:}
IRQ arrives and is injected into the VM. \mbox{Steps~4-7:} VM reprograms
timer chip for the next interrupt by issuing a hypercall.
\mbox{Step~8:} woken guest task is resumed.
}
\label{fig:timer_reprog}
\end{figure}

Our most important optimization concerned the delay incurred by
reprogramming the timer source.
Usually Linux sets a new timeout promptly after the arrival of a timer
interrupt, i.\,e. directly in interrupt context before resuming any
(possibly real-time) process. The sequence of events is shown in
Figure~\ref{fig:timer_reprog}.
Since programming the timer involves leaving and reentering the VM and
is thus a costly operation, the guest was
adapted to check whether the expired timer is about to wake a real-time
task and to postpone the timer programming until after the task's
execution in these cases.
This allowed us to move the steps 4-7 off the critical path.
Once the real-time task has completed, Karma is asked to set the next
timeout.
From the guest's perspective this
behavior is correct because the real-time load runs with the highest
priority and cannot be preempted by any other process.
If the timeout has not led to a wake-up of a real-time task, the timer
is programmed immediately as usual.
Karma takes
provisions against run-away processes by setting an emergency timeout
that is far larger than the worst-case execution time of the current
job.
Either a regular time rearming or the completion
notification of the real-time task will let Karma cancel the emergency
timeout.

Our second set of optimizations aimed at reducing the overhead caused
by lower-priority interrupts arriving during execution of a real-time
task.
As the initial version of Karma injected interrupts
into the VM in a strict FIFO manner, the first improvement was to
sort the pending interrupts and deliver the one with the highest
priority first, thereby retaining hardware semantics up to the VMM--VM
interface, and holding back the pending interrupts with lower priorities.
Once the VM reprograms the timer, Karma can be sure that the real-time
task has completed its time slice and deliver the interrupts.
For the next iteration we allowed Karma to preclude the microkernel
from delivering low-priority events to the virtual machine altogether.
The occurrence of such events still causes a VM exit and
reentry in this scenario, but the additional context switch to Karma
is inhibited. Once the real-time task has completed and the VM has announced
to Karma that event processing can resume, the microkernel delivers the
events to Karma which in turn injects them into the guest.
As a final step, we allowed Karma to
use the system's hardware task priority register (TPR) to directly inhibit the
generation of hardware interrupts below a chosen priority. This scenario
finally also saves the costly VM exits and reentries during the execution
of high-priority code by keeping interrupts pending right on the interrupt
controller.

\section{Evaluation}
\label{sect:evaluation}

To evaluate the feasibility of our design, we ran a number of
experiments. Our test machine mounted a AMD Phenom II X4 3.4GHz, 4GB
RAM (DDR3, 1333MHz) on a Gigabyte GA-770TA-UD3 board, and a Hitachi
hard disk (500GB, 7200rpm, 16MB cache). The virtual machines were
provided with 256MB RAM.

The guest running the realtime load was running linux version
3.0.14-rt31 with the \emph{PREEMPT\_RT}\cite{OSADL:RrealtimeLinux} and
Karma patches applied.  In the guest VM, we used
\emph{cyclictest}\cite{cyclictest} (version 0.83), a utility waking up
periodically and measuring the deviation from the expected wakeup
time.  These delays are a good measure for the preemptibility of the
operating system. Our version of cyclictest is slightly modified in
that it reports the measured delay directly through a hypercall to
Karma. Unless noted otherwise, the load was generated in the same VM
with ``tar -xf linux.tar.bz2'' unpacking a disk-resident archive.  For
comparision, we ran the same Linux configuration except for the Karma
patches on bare hardware. If not noted otherwise, all measurements
ran for one hour.

For our experiments, the guest running timesharing loads was granted direct
disk access, but had to use the secure GUI as console. The GUI server has to copy the
framebuffer content of a client into the device framebuffer.
Without the decoupling described in subsection \ref{subsect:VMM}, it
would not be possible to provide the user with a responsive user
interface and run real-time tasks with tight deadlines at the same
time.

Fiasco.OC offers a tracebuffer, which holds kernel generated and user
supplied logging messages. It can be made visible as read-only region
in the address spaces of user-level tasks. We made extensive use
of this facility as time-stamped log messages were often the only
way to pinpoint causes of long delays.

\newif\ifcalibration
\subsection{Base Line Measurements}

Our first measurement differs from the others in that it is the only one not
concerned with latencies.  While not the primary target, throughput is
nonetheless of interest.  Figure~\ref{fig:linuxcompile} lists the
times it takes to compile the Linux kernel. The figures are only meant
to gain an understanding of relative performance, as the imposed
limitations (one CPU, 256MB memory) certainly hurt. The result of
approximately 3.5\% slowdown is consistent with earlier
results\cite{liebergeld2010secVM}.

\begin{figure}[h]
 \begin{center}

  \begin{tabular}{|l|r|r|r|}
  \hline

   Compile time     & min & max & average \\
  \hline
  Native Linux      & 619.5s & 620.4s &  619.9s \\
  Karma Linux       & 639.2s & 641.0s &  640.3s \\
  \hline
  \end{tabular}
  \caption{Linux kernel compile benchmark, in seconds, smaller is
    better.  At least three runs were performed. All setups were
    limited to one CPU and 256MB memory. In the case of Karma, disk
    accesses were carried out directly without VMM intervention.}
  \label{fig:linuxcompile}
 \end{center}
\end{figure}

\begin{figure}
\resizebox{\columnwidth}{!}{\input{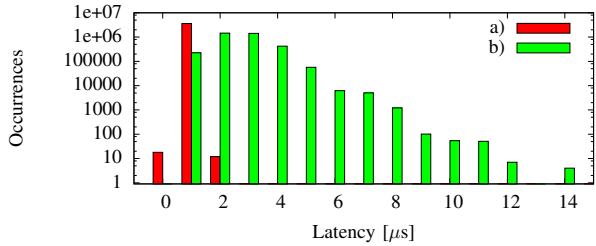}}
\caption{Native Linux with RT\_PREEMPT patch latencies measured with cyclictest (a) while
cyclictest is the only load and (b) while device interrupt inducing load is 
running. (The 0 bin denotes latencies below 1$\mu$s.)}
\label{fig:native_load_vs_idle}
\end{figure}

\begin{figure}
\resizebox{\columnwidth}{!}{\input{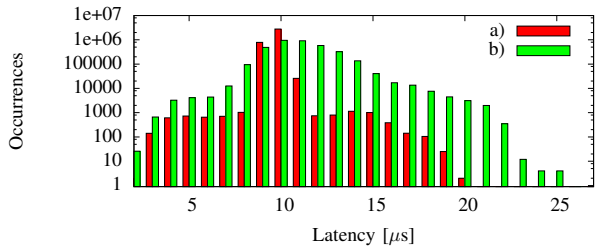}}
\caption{Baseline latencies measured with cyclictest in a guest running on a Karma
virtual machine (a) while cyclictest is the only load and (b) while device interrupt inducing load is running in the same guest.}
\label{fig:karma_load_vs_idle}
\end{figure}

In our next measurement, we ran the guest in the base configuration as
shown in figures~\ref{fig:native_load_vs_idle} and
\ref{fig:karma_load_vs_idle}. The important point to note is that the
worst-case latency under load increases less than twofold for the
virtualized case (14$\mu$s vs. 25$\mu$s).

\subsection{Bridging the Semantic Gap}

\begin{figure}
\resizebox{\columnwidth}{!}{\input{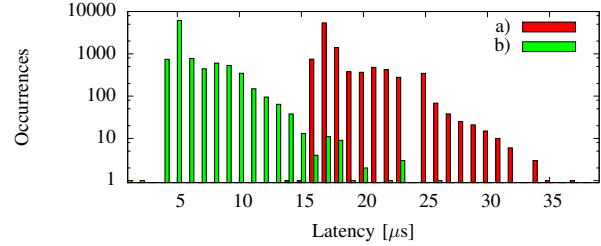}}
\caption{
(a) Baseline latencies measured with cyclictest. (b) Latencies with
deferred timer programming.
}
\label{fig:apic_timer_programming}
\end{figure}

In this chapter we demonstrate how the
real-time performance can be improved if the actions of the guest and
the layers underneath it are better coordinated. Figure
\ref{fig:apic_timer_programming} shows how the programming of the
simulated local APIC timer is rather expensive. By deferring the timer
reprogramming until the realtime workload has finished, we can improve
the release latency of the realtime workload dramatically.

\begin{figure}
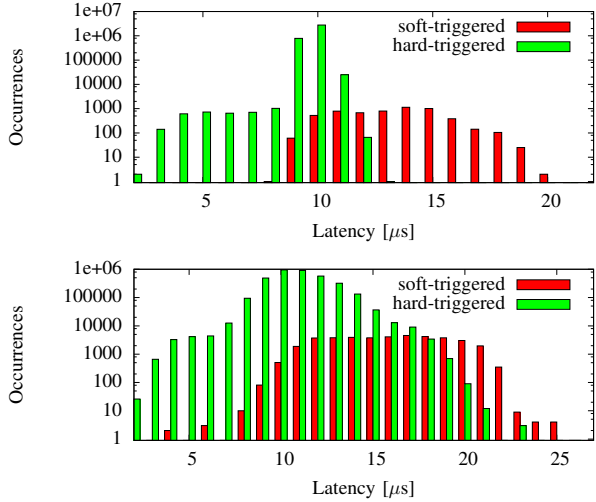

\resizebox{\columnwidth}{!}{\input{soft_vs_hard_idle.tex}}
\resizebox{\columnwidth}{!}{\input{soft_vs_hard_load.tex}}
\caption{%
Latency comparison of interrupts delivered directly from the hardware ("hard trigger")
against interrupts synthesized by Karma when the desired wakeup time has already
been reached ("soft trigger"). The figure on
top shows the latencies on an otherwise idle system, the bottom one on a system with
a device interrupt inducing load. Delayed timer programming is active in both setups.
}
\label{fig:soft_vs_hard_trig}
\end{figure}

There are however situations which can not be improved by postponing
the reprogramming. Usually Linux refrains from programming the timer if
the required delay is so small that it can not be reliably programmed; but
it does not take the expensive round-trip to the VMM into account. Thus if
Linux calls out to the VMM in order to reprogram the timer for a delay that
is above its own programming threshold but less than the actual time required
to exit and reenter the VM, Karma will refuse to program the device and
directly inject a synthetic timer event into the VM. This saves the
additional delay of programming the timer chip, but the costs incurred by
leaving and reentering the VM cannot be remedied. The latencies experienced
during such a series of events (termed a "soft trigger" as opposed to a
"hard trigger", which means an actual hardware timer interrupt) are shown
in Figure~\ref{fig:soft_vs_hard_trig}.
In the future, we will try to reliably detect close timeouts in Linux.
In such situation it might be better to poll instead of calling the
hypervisor.

\begin{figure}
\resizebox{\columnwidth}{!}{\input{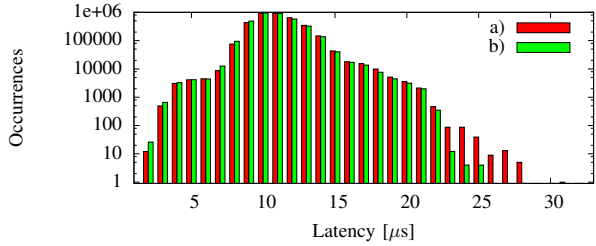}}
\caption{Latencies measured inside the virtual mashine with (b) and without (a)
optimizations in place.}
\label{fig:karma_opt_vs_unopt}
\end{figure}

Upon an HPET interrupt, Fiasco raises the TPR\footnote{The {\emph{task
      priority register}} allows to specify a threshold for interrupt
  delivery. Interrupt vectors with a numerically smaller value are not
  delivered.}  value such that only its own timer interrupts for
scheduling can pass. If the HPET interrupt results in the release of
an event handler, then it can run without interference from device
interrupts. After the handler has signaled the VMM its completion, the
latter unmasks the HPET. Fiasco also detects this and lowers the TPR
to the regular level.  Figure~\ref{fig:karma_opt_vs_unopt} shows the
the effect of that procedure. The worst case latencies grow by 6$\mu$s
if the TPR optimization is disabled.

\subsection{Isolation at VM Level}

\begin{figure}
\resizebox{\columnwidth}{!}{\input{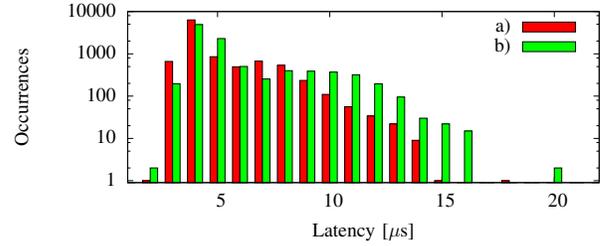}}
\caption{
Overall latencies measured by cyclictest (a) where load is induced in the same
guest as the real-time workload and (b) where load is induced in a second
virtual machine on the same physical machine.
}
\label{fig:load_in_2nd_vm}
\end{figure}

In our final measurement we investigated whether load running in a separate
VM has an impact on the real-time latencies. To that end, we ran 
our workload in a second VM with a Linux 2.6.35.2 and Karma patches applied.
Figure \ref{fig:load_in_2nd_vm} shows the meaurement results alongside measurements
with load running in the real-time VM. We put the higher latencies down to 
heavier pressure on caches and TLBs.

\section{Related Work}
\label{sect:related_work}

Linux has been used in quite a few projects to serve as the general
purpose part in a hybrid system\cite{RTLinux}\cite{DBLP:conf/edcc/CrespoRM10}\cite{1199229}.
Limited
to scheduling, the real-time layer has to rely on the Linux kernel for
services like program loading or memory access. To circumvent the
intricate user-level management of Linux, which involves paging,
real-time tasks are often implemented as kernel modules. Kernel
memory is not paged under Linux, the real-time executive can be sure
that a once-loaded task is resident. Since all real-time applications
run in the same kernel address space, faults cannot be contained with
address spaces. Although recent versions of Xenomai offer user-level
real-time applications, they still limit them to a specific API.
Maintaining more than the Linux API is seen as too cumbersome by many
developers, though.
Regardless of whether user tasks are encapsulated, the Linux kernel
runs in kernel space in any case. Its immense size makes the presence
of defects, especially in device drivers, highly likely.

Kiszka\cite{kiszka2009RTQemu} evaluates the use of Linux as a
real-time hypervisor. Using KVM and Linux-RT as a host and a guest
system latencies are evaluated. A paravirtualized scheduling interface
is introduced to improve the scheduling in the guest. The evaluations
show that reasonable scheduling latencies can be achieved.

Rehosting an operating system on top of a microkernel has been
investigated before
\cite{MkLinux}\cite{Hartig:1997:P9S:268998.266660}\cite{LeVasseur:2004:UDD:1251254.1251256}.
The approach reflects the lacking hardware virtualization support at
the time. On architectures without tagged TLBs, performance is
noticably degraded because of the context switch now necessary for a
(guest) kernel entry.  
Still, it is possible to achieve reasonable guest event-handling latencies
with such approaches\cite{rtlws2011rt_l4lx}.

Bruns at al\cite{Bruns:2010:EMV:1845876.1845967} evaluated the
application of a microkernel based system for real-time. A rehosted
RTOS was run on a commercially available embedded platform and
compared to its native version. The evaluation showed that the
benefits of consolidation outweighed the performance impact of the
virtualization approach.

EROS\cite{Shapiro99eros:a} demonstrated that a capability system can
be built on a small kernel and yield message passing performance on a
par with other then state-of-the-art microkernels. 
Unlike other systems that expose disk space through a file system,
EROS went for a persistent model. That state of the system resides
primarily on disk and is only fetched into memory for execution.  This
classical example for caching entails making decisions which objects
are kept in memory and which are written back to disk. EROS did not
provide any mechanisms to influence this caching, giving real-time
applications no chance to ensure they stay memory-resident.

Mesovirtualization, proposed by Ito and Oikawa in
\cite{DBLP:conf/seus/ItoO07}, is a virtualization approach targeted at
the x86 architecture that minimally changes the guest operating
system. As a guest they apply Linux on their VMM implementation
Gandalf, and show a better performance when comparing with Linux on
Xen, attributing their advance to their annotations.

In \cite{Ito:2008:LSP:1371608.1372768} Ito and Oikawa have implemented and evaluated shadow
page-table implementations in the Gandalf VMM. They achieve better
results than Xen, however they modify the page handling
code of their Linux guest.

Kinebuchi et al. \cite{DBLP:conf/hpcc/KinebuchiSON08} propose task
grain scheduling to host a Linux and an RTOS on an L4 based hypervisor
on embedded systems.  The guests are adapted to run in this
environment and priorities of guest tasks are mapped to host ones.
Their evaluation shows that interrupt latency is increased
significantly compared to a native RTOS.

The current excitement for virtualization obscures the fact that the
layers need to implement virtual machines may grow the attack surface,
rendering the system as whole even more vulnerable. Peter at
al\cite{peter1} have shown that kernel support can be added to
microkernels with minimal effort without undermining the security
properties of the system. This work did not seek to minimize the
trusted computing base for the VM itself, though, which was
demonstrated later \cite{Steinberg:2010:NMS:1755913.1755935}.
Liebergeld at al\cite{liebergeld2010secVM} further demonstrated that
virtualization does not hurt the real-time performance of application
running alongside the VM. The feasibility of the VM to support
real-time applications inside was not investigated.

\section{Conclusion}

In this paper, we have presented an architecture that allows for the
execution of real-time tasks in a virtual machine. Our results
indicate that real-time performance comparable to loads run on bare
hardware is only possible if the guest can give hints at its current
execution status to lower layers of the software stack.

\bibliographystyle{IEEEbib}
\bibliography{./bib,./master}

\end{document}